\documentclass[intlimits,twoside,a4paper]{article}

\usepackage[eqsecnum]{cmpj3}

\usepackage[cp1251]{inputenc}

\usepackage{bm}

\issue{2024}{27}{1}{13604}
\doinumber{10.5488/CMP.27.13604}

\title [Water-like models in contact with solid surfaces]
{Revisiting the wetting behavior of solid surfaces by water-like models
within a density functional theory}

\author[A. Kozina, M. Aguilar, O. Pizio, S. Soko{\l}owski]
{A. Kozina\orcid{0000-0002-4287-2953}\refaddr{label1},
M. Aguilar\orcid{0000-0003-3850-1188}\refaddr{label1},
O. Pizio\orcid{0000-0001-8333-4652}\refaddr{label1},
S. Soko{\l}owski\orcid{0000-0003-0580-5214}\refaddr{label2}
\thanks{Corresponding author: \email{oapizio@gmail.com}.}}

\addresses{
\addr{label1} Instituto de Qu\'{i}mica, Universidad Nacional Aut\'{o}noma de M\'{e}xico,
Circuito Exterior, 04510, Cd. de M\'{e}xico, M\'{e}xico
\addr{label2} 
Department of Theoretical Chemistry,
Maria Curie-Sk{\l}odowska University,  Lublin 20-614, Gliniana 33, Poland}


\date{Received October 16, 2023, in final form November 28, 2023}

\begin{document}
\maketitle


\begin{abstract}
We perform the analysis of  predictions of a classical density functional theory 
for associating fluids with different association strength
concerned with wetting of solid surfaces. 
The four associating sites water-like models with non-associative square-well attraction
parametrized by Clark et al. [Mol. Phys., {2006}, \textbf{104}, 3561] are considered.
The fluid-solid potential is assumed to have a 10-4-3 functional form. 
The growth of  water film on the substrate upon changing
the chemical potential is described.
The wetting and prewetting critical temperatures,
as well as the prewetting phase diagram
are evaluated for different fluid-solid attraction strength from the analysis of the
adsorption isotherms.
Moreover, the temperature dependence of the contact angle
is obtained from the Young equation. It yields estimates for the wetting temperature as well.
Theoretical findings are compared with experimental results and in a few cases with data from computer
simulations. The theory is successful and quite accurate in describing the wetting temperature
and contact angle changes with temperature for different values of fluid-substrate attraction. 
Moreover, the method
provides an easy tool 
to study other associating fluids on solids of importance 
for chemical engineering,
in comparison with laboratory experiments and computer simulations.

\printkeywords


\end{abstract}
\section{Introduction}

This paper is dedicated to our good friend for many years, Dr. Jaroslav Ilnytskyi, 
on the occasion of his 60th birthday. 
Dr. Ilnytskyi made several important contributions along different lines of research within
the theory and computer simulations of liquids and solutions involving complex
molecules and of fluid-solid interfacial phenomena~\cite{slavko1,slavko2,slavko3}.
We have been honored and benefited from the ideas and participation of 
J. Ilnytskyi in our common projects during last decades~\cite{slavko4,slavko5,slavko6,slavko7}.

Wetting is one of the most important phenomena occurring at 
the liquid-solid interface that determines almost 
an endless number of practical applications. An adequate 
description of wetting in natural and  synthetic systems involving 
fluid-solid interfaces represents a challenging
subject for theoreticians.
Principal aspects in 
theoretical description of wetting 
and its consequences for various applications 
were described in several monographs and reviews, see e.g., \cite{i1,i2,i3,i4}.
Early theoretical works restricted to models of geometrically smooth, planar, 
and energetically homogeneous solid substrates,
so that each element of the solid surface exerts the same action on the adjacent fluid phase.
More recent extensions of the theory were concerned with the wetting of heterogeneous solids by fluids
\cite{i5,i6,i7,i8,i9,i10}.
Water is one of the most important chemical compounds in nature. Hence,
the phenomenon of wetting the surfaces by water is crucial in many processes of everyday life~\cite{i4}.
It is therefore not surprising that the problem are extensively studied
in many works, some of which were quoted in our previous recent contributions~\cite{MolPhys,encyclopedia}.

The application and the development of theoretical approaches, as well as the use of
computer simulations in wetting studies require
the knowledge of interaction potentials. The models for describing water-water
interactions used in computer simulations are much more sophisticated than 
those used in theory. Their appropriateness with respect to
experimental observations can be verified either by comparing the theoretically
predicted and experimentally determined microscopic structure
of bulk water~\cite{pusztai} or thermodynamic properties.
If thermodynamic aspects are the principal focus, theoretical approaches
are commonly based on the versatile and 
technically convenient SAFT (Statistical Associating Fluid Theory) 
methodology~\cite{Saft1,Saft2}.

The SAFT is a perturbation-type method,
 according to which the intermolecular interaction
 between water molecules is commonly approximated 
by a hard-core repulsion, short-range attraction 
and orientation-dependent associative potential to mimic hydrogen bonding.
More sophisticated versions of the method take into account the dipole-dipole
intermolecular interaction as well \cite{mccabe}. Obviously, the 
dielectric properties of water are out of question within simplified water-like models.

The  SAFT approach that takes into account the hard-core repulsion, the non-associative 
square-well or Lennard-Jones attraction and the associative potential
was implemented in density functional theories (DFTs)  of nonuniform fluids, 
see e.g.,~\cite{DF} for a comprehensive description of the methodology.
The developed methods~\cite{DF0,DDF0,DF1,DF2} are
principally used in the studies of surface phase transitions,
such as capillary condensation and wetting transitions.
In particular, it was shown in~\cite{MolPhys} that the SAFT-type  model
of Clark et al.~\cite{Saft2} with square-well
nonassociative attraction can appropriately describe
the temperature dependence of the gas-liquid
surface tension of water and the wetting transition of water on
graphite-like surfaces in agreement with experimental data.
A more recent parametrization 
of water-like  models~\cite{jackson}, with attraction described by Mie potential, has not
been implemented in the DFTs for inhomogeneous fluids so far.

The aim of this work is manyfold. We would like to  briefly review the
principal elements of the theory needed to characterize the
wettability of a solid surface by associating fluids and by
water as an example. Critical comments are provided concerning the
application of this theoretical construction.
On the other hand, we present novel elements 
and capture types of the first-order surface phase
transitions predicted by phenomenological approach
of Pandit et al.~\cite{pandit82} used in \cite{hughes2014} for lattice model fluids.
In this respect, our contribution is extension of the recent study \cite{MolPhys}
focused on a single water-like model with dominating associative
inter-particle interaction within SAFT-type approach.
The surface phase diagrams are evaluated from the analysis
of adsorption isotherms of water-like models. The present study
is carried out for the bulk densities up to the bulk gas-liquid coexistence.

The paper is arranged as follows.
The models for water-water and water-surface interactions are discussed in the next section.
Next, we briefly outline the theory for a bulk system and the method of
evaluation of the parameters of water-water potential.
Then, we recall the elements of the DFT  used in the studies of nonuniform fluids.
In the ``Results and discussion'' section, divided into two subsections
we first focus on the presentation of bulk liquid-vapor phase diagrams 
and the results for the temperature dependence of the surface tension.
Next, we present a comparison of the results of contact angle calculations
and the surface phase diagrams for particular water-water potential models.
These results are
supplemented by presentation of examples of adsorption isotherms
and the density profiles of
fluid species.
Our principal findings  are summarized in the
last section.


\section{Modelling of associating fluids and water in contact with solid surface}

The interaction potentials between water molecules and water molecules with the solid surface used in this 
work are identical to those used in our previous publications \cite{MolPhys,encyclopedia}.
Namely, the interaction between two water
molecules is described by using the model introduced in \cite{jackson1} with the 
parameters established by Clark et al. \cite{Saft2}.
The latter publication provides the statement of reasons for the choice 
of the interaction potential form and its parameters
from experimental data for the bulk liquid-vapor coexistence of water.

Each fluid molecule possesses four associative sites denoted as A, B, C, and D
inscribed into a spherical core. However, only the site-site association
AC,  BC, AD, and BD is allowed and the set of these sites is denoted as $\Gamma$.
Thus, the pair intermolecular potential between molecules 1 and 2  depends
on the center-to-center distance,  $r_{12}=|\mathbf{r}_{12}|$,  and on molecular orientations,
$\boldsymbol{\omega}_1$ and $\boldsymbol{\omega}_2$
\begin{equation}
 u(r_{12}) = u_{ff}(r_{12}) + \sum_{\alpha,\beta \in \Gamma}  
u_{\alpha\beta}(\mathbf{r}_{\alpha\beta}),
\end{equation}
where
$\mathbf{r}_{\alpha\beta}=\mathbf{r}_{12}+\mathbf{d}_{\alpha}(\boldsymbol{\omega}_1)-\mathbf{d}_{\beta}(\boldsymbol{\omega}_2)$
is the vector connecting the site $\alpha$ on molecule 1 with the site $\beta$ on molecule 2,
 $\mathbf{d}_{\alpha}$
is the vector from the molecular center to the site $\alpha$.
The length of the vector $\mathbf{d}_{\alpha}$ is assumed constant, $d_s = |\mathbf{d}_{\alpha}|$.
The association potential is taken as
\begin{equation}
u_{\alpha\beta}(\mathbf{r}_{\alpha\beta})=
\left\{
\begin{array}{ll}
-\varepsilon_{\text{as}}, &  0< |\mathbf{r}_{\alpha\beta}| \leqslant r_c ,\\
 0,  &    |\mathbf{r}_{\alpha\beta}|   > r_c,
\end{array} \right.
\label{eq:asw}
\end{equation}
where $\varepsilon_{\text{as}}$ is the depth  and
$r_c$ is the cut-off of the associative interaction.
The non-associative part of the pair potential, $u_{\text{ff}}(r)$, is considered as
the sum of hard-sphere (hs), repulsive term and attractive, square-well (att) 
contribution
\begin{equation}
u_{\text{ff}}(r) = u_{\text{hs},\text{ff}}(r) +  u_{\text{att},\text{ff}}(r).
 \label{eq:sw}
\end{equation}
The $\text{hs}$ term is
\begin{equation}
u_{\text{hs},\text{ff}}(r) =
\left\{
\begin{array}{ll}
\infty, &   r < \sigma,\\
0,   &   r \geqslant \sigma,
\end{array}
\right.
\label{eq:hs}
\end{equation}
whereas the attractive interaction is
\begin{equation}
\label{eq:uSW}
 u_{\text{att},\text{ff}}(r)=
\left\{
\begin{array}{ll}
0, &  r < \sigma,\\
-\varepsilon, &  \sigma \leqslant r < \lambda \sigma,  \\ 
0, &  r \geqslant \lambda \sigma.
\end{array}
\right.
\end{equation}
In the above,
 $\varepsilon$ and $\lambda$ are the depth and the range of the
attraction, respectively, and $\sigma$ is the $\text{hs}$ diameter
of the spherical core. This kind of splitting of inter-particle
interaction into two terms corresponds to the Barker-Henderson type of
perturbation theory~\cite{BH2}.

The interaction of a water molecule with graphite-like solids
can be described by the potential of Steele \cite{steele}
\begin{equation}\label{steele}
 v_{sf}(z) =2\piup\rho_g\varepsilon_{sf}\sigma^{2}_{sf}\Delta
 \left[ \frac{2}{5} \left( \frac{ \sigma_{sf}}{z}\right)^{10} 
 - \left( \frac{ \sigma_{sf}}{z}\right)^{4} \right. 
\left. - \frac{\sigma_{sf}^4 }
{3 \Delta (z+0.61 \Delta)^3 }
 \right],
\end{equation}
where $\varepsilon_{sf}$, $\sigma_{sf}$ are
the energy and the distance parameters, respectively, $\Delta$ is the interlayer spacing
of the graphite planes, $\Delta= 0.335$~nm and $\rho_g$ is the density of graphite,
$\rho_g= 114$ nm$^{-3}$. The value of  $\sigma_{sf}$  results from
the combination rule, $\sigma_{sf} = (\sigma_g +\sigma)/2$,
where $\sigma_g=0.34$~nm is the diameter of carbon atoms in graphite.
Discussion of the applicability of equation~(\ref{steele}) to associative fluid-solid
surface interfaces is given
in~\cite{MolPhys}.  In particular,
this potential was successful in describing the temperature dependence
of the contact angle of water on graphite.

\section{Theory}

Our interest is in studying the gas-liquid interface and the 
behavior of  water at graphite-like surface.
This study is carried out for the water-water interaction model that was
 proposed by Clark et  al.\cite{Saft2}.

\subsection{Bulk fluid}

For the sake of completeness of description of the necessary elements
in the present work, let us briefly recall the method used to determine the set of 
five parameters of the potentials given by equations~(\ref{eq:asw}) and (\ref{eq:uSW}).
In general terms, the method was based on the SAFT theory, according to which
the free energy of the bulk uniform fluid, $F$, is considered as a sum
of an ideal, $F_{\text{id}}$ and excess, $F_{\text{ex}}$ terms.

The 
ideal term is exact. Its configurational part (i.e., the contribution apart from the kinetic term) is 
$F_{\text{id}}/VkT=\rho_\text{b}(\ln\rho_\text{b}-\rho_\text{b})$, where $V$ is the volume and $\rho_\text{b}$ is 
the fluid density. The excess contribution
is considered in a perturbational manner with respect to
the reference system of hard spheres. It consists
of the sum of terms resulting from attractive, non-associative
interaction and from association
\begin{equation}
F_{\text{ex}}=F_{\text{non}}+F_{\text{as}}.
\end{equation}

The association contribution is described at the level of  the first-order thermodynamic
theory of Wertheim~\cite{wertheim1,wertheim2} 
and is written in terms of the fraction of
molecules not bonded at a site $i$, $\chi_i$
\begin{equation}
  \frac{F_{\text{as}}}{VkT} = \rho_\text{b}\sum_{i=1}^{4} \bigl[\ln \chi_i - \frac{1}{2}(\chi_i -1)\bigr].
\end{equation}

Since all associating sites are identical according to the model in question, 
the fraction of molecules not bonded at a site $i$ is obtained from 
the single equation, which is the statistical mechanics expression 
for the mass action law
\begin{equation}
 \chi_i= \frac{1}{1+2\rho_\text{b}\chi_i\Delta_{\text{as}}}.
\end{equation}
The quantity $\Delta_{\text{as}}$ invokes the contact
value of the pair distribution function of the
reference fluid, $g_{\text{hs}}(r=\sigma)$, and  the associative Mayer function  
$F_{\text{as}}=\exp[\varepsilon_{\text{as}}/kT] -1$,
\begin{equation}
 \Delta_{\text{as}}=g_{\text{hs}}(r=\sigma)F_{\text{as}}K_{\text{as}},
\end{equation}
where $K_{\text{as}}$ is the site bonding volume. The bonding volume 
follows from the geometry of associative interaction and yields
\begin{eqnarray}
K_{\text{as}}&=&\sigma^2 [\ln\{(r_c+2d_s)/\sigma\}(6r_c^3+18r_c d_s-24d_s^3)+(r_c+2d_s-\sigma)\nonumber\\
&&\times (22d_s^2-5r_cd_s-7d_s\sigma-8r_c^2+r_c\sigma+\sigma^2)]/(72d_s^2).
 \label{eq:kas}
\end{eqnarray}
All the details concerning calculations of $\chi_i$ 
and $K_{\text{as}}$ can be found in~\cite{Saft2,jackson1,jackson2},
which are not given here to avoid unnecessary repetition. 

The non-associative contribution includes hard sphere reference and dispersion interaction
contributions according to the
representation of the potential in equation~(\ref{eq:sw}), $F_{\text{non}}=F_{\text{hs}}+F_{\text{att}}$. 
The hard-sphere term is accurately described by using the Carnahan-Starling equation of state 
\begin{equation}
F_{\text{hs}}/VkT=kT \frac{4\eta-3\eta^2}{(1-\eta)^2},
\end{equation}
where $\eta=\piup\sigma^3\rho_\text{b}/6$ is the packing fraction. 
However, the attractive interactions effect is taken into account within
the second order Barker-Henderson high-temperature
perturbation expansion with respect to the hard-sphere reference system~\cite{BH2}
\begin{equation}
F_{\text{att}}=F_{\text{att},1}+F_{\text{att},2}. 
\label{eq:expan}
\end{equation}
The first-order term, $F_{\text{att},1}$, is easier to calculate, whereas
the second-order contribution, $F_{\text{att},2}$, requires a more sophisticated
manipulation \cite{Saft2,gil}. The resulting expressions are a bit cumbersome 
because they contain a set of adjustable parameters recommended to reach
a reasonable precision of thermodynamic properties~\cite{gil}.
The free energy leads to a grand thermodynamic potential of bulk fluid
\begin{equation}
\Omega_\text{b}/V = F/V -\mu \rho_\text{b}, 
\end{equation}
or equivalently to pressure, $p=-\Omega_b$/V. 
Here, $\mu$ is the configurational chemical potential of the fluid at density
$\rho_\text{b}$ and temperature $T$.

In order to obtain
five  parameters of  equations~(\ref{eq:asw}) and (\ref{eq:uSW})
namely $\sigma$, $\varepsilon$, $\lambda$, $\varepsilon_{\text{as}}$
and $r_c$, Clark et al. \cite{Saft2} performed fitting of the
theoretical predictions for vapor pressure and saturated liquid density
to the experimental data for vapor-liquid coexistence.
The fitting procedure was carried out at various temperatures
in the interval from the triple point temperature
up to the $0.9T_\text{c}$, where $T_\text{c}$ is the bulk critical temperature. 
The numerical procedure leads to four different sets of
parameters in equations~ (\ref{eq:asw}) and (\ref{eq:uSW}) that yield
a rather accurate description of the vapor-liquid equilibrium. 
The models are designated as W1, W2, W3 and W4, see table~\ref{tab:1}~\cite{Saft2}. 
These sets can be classified 
according to the respective weight of the effects coming from attraction and association.
The models W1 and W2 
can be referred to as the models with low-dispersion and high hydrogen bonding effects,  
whereas the models W3 and W4 --- as the models with high-dispersion and low hydrogen 
bonding effects. It is worth to note that the W1 model somewhat better 
reproduces the bulk phase diagram of water than the remaining ones.

\begin{table}[h]
  \centering
   \caption{
   Four optimal sets of parameters for water-like model fluids~\cite{Saft2}.
   }
   \vspace{0.3cm}
   \begin{tabular}{cccccc
   }
  \hline \hline
 Model &  $\sigma  $ (nm)& $\varepsilon/k$ ($K$) & $\lambda$&  $r_c $ (nm)&
     \vspace{0.1cm} $\varepsilon_{\text{as}}/k$ ($K$)  \\[0.5ex]
     \hline
W1 & 0.303420 & 250.000 & 1.78890 & 0.210822 & 1400.00 \\
W2 & 0.303326 & 300.433 & 1.71825 & 0.207572 & 1336.95 \\
W3 & 0.307025 & 440.000 & 1.51103 & 0.209218 & 1225.00 \\
W4 & 0.313562 & 590.000 & 1.37669 & 0.215808 & 1000.00 \\
\hline \hline
\end{tabular}
\label{tab:1}
\end{table}

In summary, the second-order
perturbation theory was used for attractive, non-associative free energy contribution.
However, in the case of studies of nonuniform fluid systems, 
such an approach is computationally expensive, and, therefore, a 
certain simplification would be desirable.
A simpler version of the theory can rely on the substitution
of the perturbation expansion of equation~(\ref{eq:expan}) by
the expression resulting from the 
analytic solution of the first-order mean spherical (FMSA) approximation. 
This route was explored in detail for W1 and W2 water-like models~\cite{trejos1}.
It is documented that the accuracy of the approach is sensitive to the parameters 
of the non-associative interaction potential.
Moreover, a much simpler mean-field type approximation for the square-well
fluid,
\begin{equation}
 F_{\text{att}}/VkT= -4\eta \rho \varepsilon (\lambda^3-1),
\end{equation}
yields a similar accuracy of the description of vapor-liquid bulk coexistence~\cite{trejos1}.
The mean-field approximation is particularly useful in describing the adsorption of water on solid surfaces 
and on even more complex substrates formed by solids with, e.g., grafted 
chains \cite{henderfest,holovkofest}.

\subsection{Density functional approach}

Several  studies of adsorption of water on graphite-like surfaces are based 
on a version of density functional theory.
Within DFT, the equilibrium local density of water-like molecules, $\rho(\mathbf{r})$, and 
then all
thermodynamic functions are determined by minimizing
the grand potential functional \cite{DF}
\begin{equation}
\label{eq:omega}
 \Omega[\rho(\mathbf{r})] =F[\rho(\mathbf{r})]+\int \rd\mathbf{r}\rho(\mathbf{r})[v_{sf}-\mu],
\end{equation}
where $\mu$ is the chemical potential of bulk fluid at the bulk density $\rho_\text{b}$ and the 
temperature $T$. The minimization condition reads
\begin{equation}
 \frac{\delta \Omega[\rho(\mathbf{r})]}{\delta \rho(\mathbf{r})}=0.
\end{equation}

Following the theory for bulk fluids the free energy functional, $F[\rho(\mathbf{r})]$
is assumed to be the sum of the ideal and the excess parts arising from hard-sphere, non-associative
attractive forces and from chemical association. The exact expression for the 
nonuniform system configurational ideal free energy
is \cite{DF}
\begin{equation}
 F_{\text{id}}/kT=\int \rd\mathbf{r} \rho(\mathbf{r}) [\ln\rho(\rho(\mathbf{r})) -1].
\end{equation}

The evaluation of the free energy functionals resulting from
the volume exclusion effects,
$F_{\text{hs}} [\rho]$, and from associative interactions, $F_{\text{as}}[\rho]$, requires  the knowledge of 
three scalar and two vectorial  weighted local densities, $n_i$, $i = 0, 1, 2, 3$
and $\mathbf{n}_j$, $j=V1, V2$. 
The averaged densities are related to the density profile of 
particles, $\rho(\mathbf{r})$, \cite{c41},
\begin{equation}
 n_i(\mathbf{r})=\int \rd\mathbf{r'}\rho(\mathbf{r'}) w_i(|\mathbf{r'}-\mathbf{r}|),
\end{equation}
and
\begin{equation}
 \mathbf{n}_j(\mathbf{r})=\int \rd\mathbf{r'}\rho(\mathbf{r'}) \mathbf{w}_j(|\mathbf{r'}-\mathbf{r}|),
\end{equation}
where    $w_i(|\mathbf{r'}-\mathbf{r}|)$ and  $\mathbf{w}_j(|\mathbf{r'}-\mathbf{r}|)$
are the scalar and vector weight functions,
see  \cite{c41}. 
Since the equations defining the weight functions, as well as the equations
for the hard-sphere and associative free energy contributions are well-known and are
 given in previous works~\cite{c41,trejos2,trejos3}, we omit them here 
 to avoid unnecessary repetitions. We only note that for the bulk uniform system,
 they reduce to the equations presented in the previous subsection.
Finally, the mean-field approximation for the attractive, non-associative
 forces reads \cite{DF}
 \begin{equation}
 F_{\text{att}}[\rho]=\frac{1}{2}\int \rd\mathbf{r_1}\rd\mathbf{r_2}\rho(\mathbf{r}_1)\rho(\mathbf{r}_2)
 u_{\text{att},\text{ff}}(|\mathbf{r}_1-\mathbf{r}_2|).
 \label{eq:mf}
\end{equation}

In the studies of interfacial phenomena, the bulk fluid in contact
with a solid can be either in gaseous or in a liquid state. If the bulk
density is equal to the density of gaseous, $\rho_\text{b} = \rho_\text{bg}$, or liquid water,
$\rho_\text{b}=\rho_\text{bl}$, at the gas-liquid coexistence,
then the symbols $\Omega_\text{sg}$ and $\Omega_\text{sl}$ refer
to the excess grand potentials for the bulk gas or liquid coexisting phases
in contact with a surface. They are calculated as the left-hand side or
right-hand side limits
$$
 \Omega_\text{sg}=\lim_{\rho_\text{b}\to \rho_\text{bg}^+} \Omega(\rho_\text{b}), 
$$ 
\begin{equation}
 \Omega_\text{sl}=\lim_{\rho_\text{b}\to \rho_\text{bl}^-} \Omega(\rho_\text{b}).
\end{equation}

Within the  DFT presented above, we can also obtain the
values of the surface tension, $\gamma$.
To determine the surface tension, $\gamma$,
one needs to calculate the density profile
across the interface between two coexisting liquid and gaseous phases. 
For this purpose, we
remove the solid (i.e, we remove the  potential $v_{sf}$ from 
equation~(\ref{eq:omega}) and set the boundary conditions
 $\rho(z = - \infty) = \rho_\text{bl}$ and 
$\rho(z = \infty) = \rho_\text{bg}$.
The grand canonical thermodynamic potential for such a system is $\Omega_\text{lg}$.
Then, the surface tension is
\begin{equation}
 A\gamma=\Omega_\text{lg} - \Omega_\text{b},
 \label{eq:gamma}
\end{equation}
where $A$ is the surface area of the gas-liquid interface and the bulk quantity
$\Omega_\text{b}$ is obtained for gaseous or for liquid coexisting bulk density (the condition
of mechanical equilibrium imposes their equality).
The details of the calculations were 
presented in~\cite{bryk004}.

Finally, we would like to comment on the application of the DFT
in calculations of the liquid-solid contact angle, $\theta$. 
The contact angle is the angle between the liquid  in equilibrium
with a gas and a solid surface where they meet.
More precisely, it is the angle between the 
surface tangent on the liquid-vapor interface and the surface tangent on the solid-liquid 
interface at their intersection, measured throughout the liquid phase.
The value of the contact angle is commonly used to characterize the wettability of a solid surface.
For a given system consisting of solid, liquid, and vapor at a given temperature,  
the equilibrium contact angle $\theta$ possesses a unique value. 
In experimental works, the value $\theta=90^\circ$
is distinguished as delimiting the regimes of convex and concave liquid menisci
in capillary phenomena.

Theoretical description of the contact angle results from  
thermodynamic equilibrium between the three coexisting phases: 
 the solid phase, and the liquid and  gas phases and
 can be is related to the  gas-solid, $\Omega_\text{sg}$,
 gas-liquid, $\Omega_\text{sl}$, and the gas-liquid, $\gamma$, grand potentials
 via the Young equation \cite{Dietrich88},  
\begin{equation}
 \Omega_\text{sg}-\Omega_\text{sl}=A\gamma \cos\theta.
 \label{young}
\end{equation}

The
Young equation is an approximation to reality \cite{78,79}
as it neglects the effect of the line tension at the three-phase
contact. 

\section{Results and discussion}

Following our previous works \cite{MolPhys,encyclopedia,trejos1,trejos2,trejos3},
we define the dimensionless parameters
describing the interactions in the system by using $\sigma$ and $\varepsilon$ of 
equation~(\ref{eq:uSW}) as units of length and energy, respectively.
We have then, $\varepsilon_{\text{as}}^* =\varepsilon_{\text{as}}/\varepsilon$, and
$r_c^*=r_c/\sigma$. For the sake of convenience, we  also use the
notation
\begin{equation}
\varepsilon_{gs}^*=2\piup\rho_s\varepsilon_{sf}\sigma_{sf}^2\Delta/\varepsilon
\label{eq:egs}
\end{equation}
for the energy parameter in equation~(\ref{steele}).
The reduced thermodynamic parameters are:
$T^*=kT/\varepsilon$,  $\mu^*=\mu/\varepsilon$ and $\rho_\text{b}^*=\rho_\text{b}\sigma^3$,
in close similarity to the reduced units in Lennard-Jones systems, see e.g.,~\cite{rap}.

Table \ref{tab:2} contains the reduced values of
the water-water interaction for the water-like models in question. 
In all cases, the
center of the molecule -- association site distance is $d_s^*=d_s/\sigma=0.25$.
Moreover, table \ref{tab:2} also provides the values for the bonding volume $K^*=K_{\text{as}}/\sigma^3$ and for
the critical temperatures that result from the mean-field version of the theoretical procedure for
each of the bulk models \cite{trejos1}.

\begin{table}[h]
  \centering
   \caption{
   Dimensionless parameters for the W1, W2, W3 and W4 four-site water-like models.
}
   \vspace{0.2cm}
   \begin{tabular}{ccccccc
   }
  \hline \hline
 Model \ & $\varepsilon_{\text{as}}^*$ & $K^*$ & $r_\text{c}^*$ & $T^*_\text{c}$ \\
     \hline
W1 & 5.600  & 0.03820 & 0.6948 & 2.718 \\
W2 & 4.4501 & 0.03202 & 0.6843 & 2.193 \\
W3 & 2.7841 & 0.03046 & 0.6814 & 1.330 \\
W4 & 1.6949 & 0.03424 & 0.6882 & 0.862 \\
\hline \hline
\end{tabular}
\label{tab:2}
\end{table}

The basic difference between 
common Lennard-Jones reduced units \cite{rap}
and those used in this work should be emphasized. In the former case, 
the value of $\varepsilon$
used for defining the reduced temperature is
the minimum of the attraction between a pair of molecules. For associating
fluids, the value of $\varepsilon$ results from non-associative forces only,
but the total effective attraction between a pair of particles is much
higher, due to site-site association. Therefore, the values of the reduced critical
temperatures from table \ref{tab:2} differ very much from the critical
temperature of a Lennard-Jones system \cite{hey}.


\subsection{Bulk gas-liquid coexistence}

As we have already mentioned, the most simple, mean field
approximation is commonly used within a DFT for nonuniform systems.
Application of the theory of non-homogeneous systems must be preceded by an 
examination of its assumptions in the case of homogeneous systems.
Therefore, we check how the introduced assumptions affect the
phase diagram. The accuracy of the mean-field theory
for bulk phase diagram predictions was studied in detail in~\cite{trejos1,yukal}.

\begin{figure}[h]
\centering{
\includegraphics[height=6.0cm]{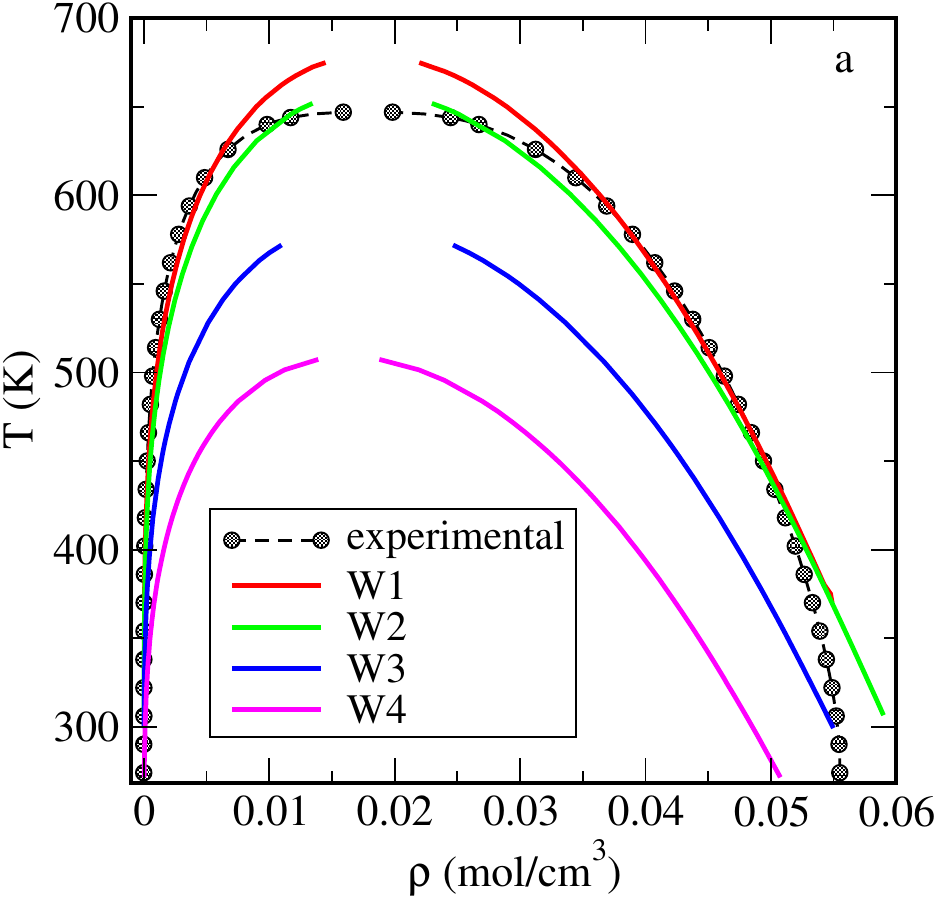}
\includegraphics[height=6.0cm]{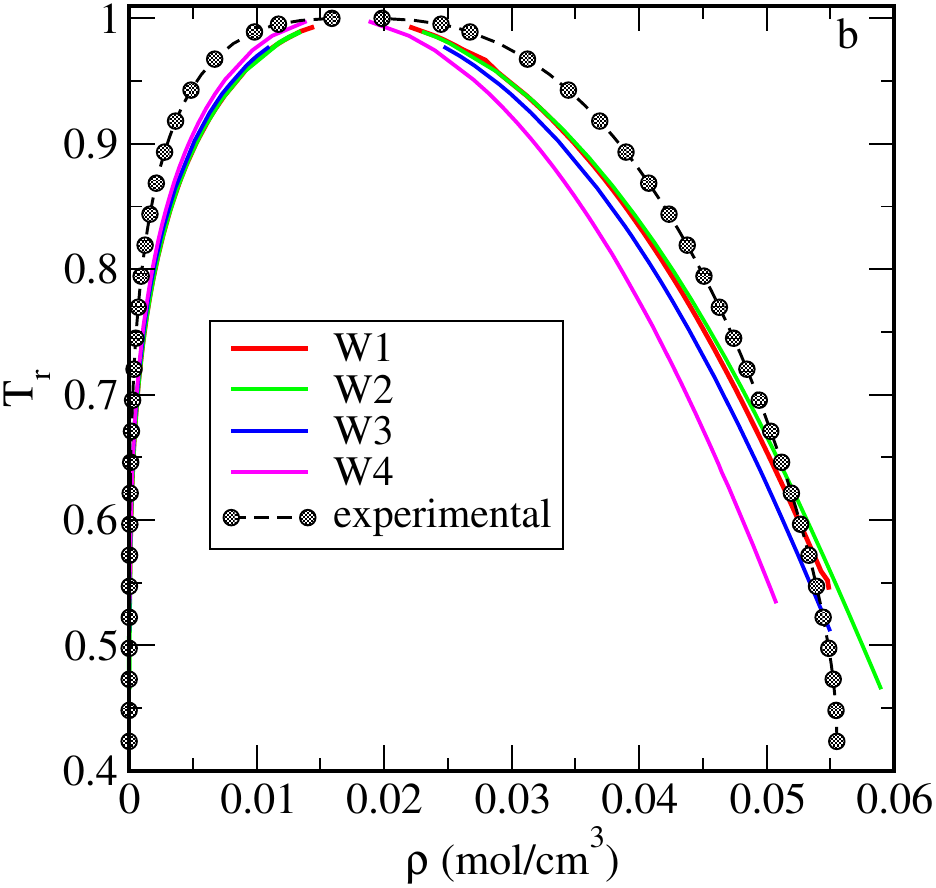}
}
\caption{(Colour online) 
A comparison of the gas-liquid phase diagrams 
in the density-temperature plane
for the W1, W2, W3 and W4  models of water. Panel a is in real
units. In panel b, the temperature is rescaled using
the bulk critical temperature for each model, $T_\text{r}=T/T_\text{c}$, cf. table~\ref{tab:2}.
The experimental data (symbols) are from \cite{diagram,expe}. 
}
\label{fig:1}
\end{figure}

In the left-hand panel of figure~\ref{fig:1} we display the 
bulk phase diagram in real temperature and density units. 
As we see, a reasonable agreement is observed
for the models W1 and W2 with  low-dispersion and high hydrogen bonding contributions
to the pair energy. Two remaining models, W3 and W4 lead to much worse
predictions of the critical temperature, measured in real units. Note that the accuracy
of the critical densities is similar for all the models.
However, the predictions of the phase diagram 
for W3 and W4
models look better, if the temperature is rescaled by using the
bulk critical temperature, $T_\text{r}=T^*/T_\text{c}^*$,
see the right-hand panel of figure~\ref{fig:1}. 

\begin{figure}[h]
\centering{
\includegraphics[height=5.5cm,clip]{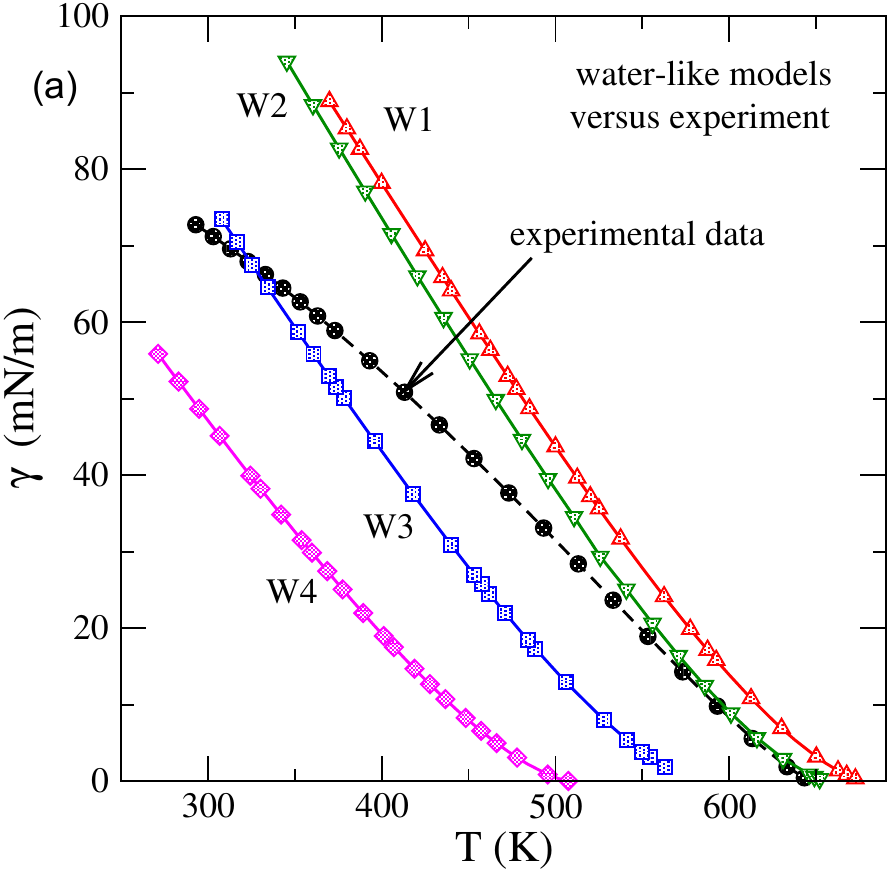}
\includegraphics[height=5.5cm,clip]{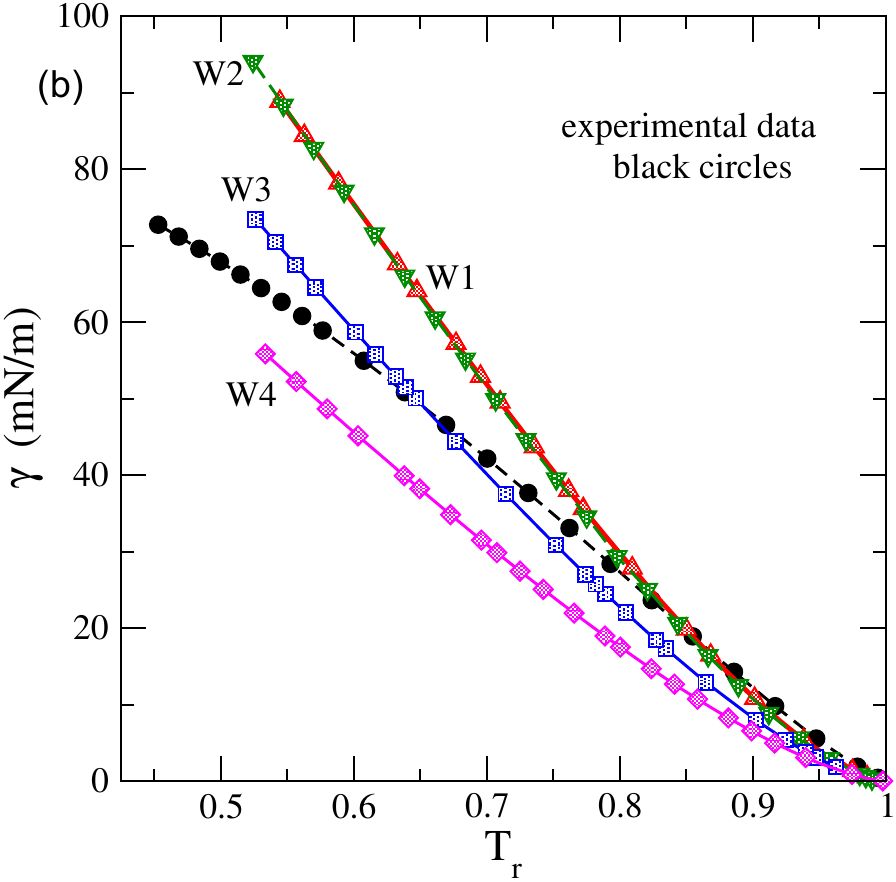}
}
\caption{(Colour online) 
A comparison of the temperature dependence of the surface tension of water
from theory for the models W1, W2, W3 and W4 (lines) with experimental data (symbols) \cite{expe}. 
Panel a is in real units, while in panel b  the temperature
is rescaled by the bulk critical temperature, cf. figure~\ref{fig:1}.
}
\label{fig:2}
\end{figure}

Figure \ref{fig:1}b indicates that the simple mean field approximation used
within the SAFT approach is capable of predicting the phase
behavior of bulk with reasonable accuracy, if the temperature is appropriately rescaled 
(note that the temperature rescaling is equivalent with rescaling
of the energy parameter $\varepsilon$). Similarly to figure~\ref{fig:1}a,
the results in figure~\ref{fig:1}b
indicate a slight superiority of the models W1 and W2,
i.e., the models with  low-dispersion and high hydrogen bonding 
compared to the models with  high-dispersion and low hydrogen bonding effects.
One important drawback of the description of association effects
can be seen in figure~1. Namely, the behavior of the liquid branch at low temperatures
is not well captured within the simplified theoretical description here, as well
as in the original modelling~\cite{Saft2}. This precludes the description of 
density anomaly of water close to the experimental triple point as well as
of the anomaly of isothermal compressibility. These features are out of reach within the present approach,
in contrast to water models used in computer simulations~\cite{vega}.

For each potential model, the obtained  mean field values for the  
coexisting dew and bubble densities,
 $\rho_\text{bg}$ and
$\rho_\text{bl}$,  are next used for evaluating the density profiles
across the gas-liquid interface at several temperatures and then --- the
values for the surface tension, $\gamma$ (see figure~\ref{fig:2}).

In figure~\ref{fig:2}a, real units are used. However, the quantities from DFT
are expressed in reduced, dimensionless units. Therefore,
a comment is necessary regarding the conversion of theoretically determined quantities into real
units.  According to equation~(\ref{eq:gamma}), the surface tension is
defined as the surface excess grand thermodynamic potential per unit surface area and
the computed values of the surface tension  are dimensionless,
$\gamma^*=\gamma\sigma^2/\varepsilon$.  To obtain the values of $\gamma$
in mili Newtons per meter, for each model we use the appropriate molecular 
parameters from table~\ref{tab:1}.
On the other hand, due to the difference of the critical temperatures
predicted for the different models in question (see table~\ref{tab:2}),
the surface tension curves in figure~\ref{fig:2}a end up at different  temperatures. 
Thus, similarly to figure~\ref{fig:1}b,  in figure~\ref{fig:2}b,
we use the rescaled temperature units $T_\text{r}=T^*/T^*_\text{c}$. However, for the values
of the surface tension we keep the real units. 
This presentation differs from 
our previous work \cite{MolPhys}. 
The data displayed in figure~\ref{fig:2}b
indicate that for the models W1 and W2, the theoretical values
overestimate the surface tension in comparison with experimental data 
(especially at lower temperatures $T_\text{r}$)
while the theoretical data for W3 and W4 models underestimate the surface tension.

The agreement of theoretically predicted values of surface tension with experimental
data \cite{expe} is not as good as from calculations of Clark et al. \cite{Saft2}.
However, the calculations of the surface tension of
 Clark et al.\cite{Saft2} within the second-order perturbation expansion for the
attractive forces follow from a version of the SAFT-DFT by Gloor et al.~\cite{gloor}
tested solely for vapor-liquid interfaces. 
Our data in figure~\ref{fig:2} are obtained within the mean field
approximation.  Moreover, our calculations lead to bigger discrepancies
between the values of $\gamma$ for different water-water interaction models in comparison 
with~\cite{Saft2}. Apparently, there remains room for optimization of the parameters 
of different contributions to the inter-molecular interaction potential 
within mean field and higher-order approaches to improve the description of
the particular property of interest. Cancellation of errors from applied
approximations for a given property does not ensure the applicability of the
theory in a wider context. Therefore, we proceed now to a more demanding
test for the theory by considering water-solid interface.

\subsection{Water in contact with solid surfaces. Contact angles}

In this subsection of our work we focus on the problem of the description
of wetting of solid surfaces for different models of water-water interactions
in terms of contact angles.
We begin our discussion with the presentation of the temperature dependence
of the contact angle, calculated from equation~(\ref{young}).

\begin{figure}[h]
\begin{center}
\includegraphics[height=6.0cm]{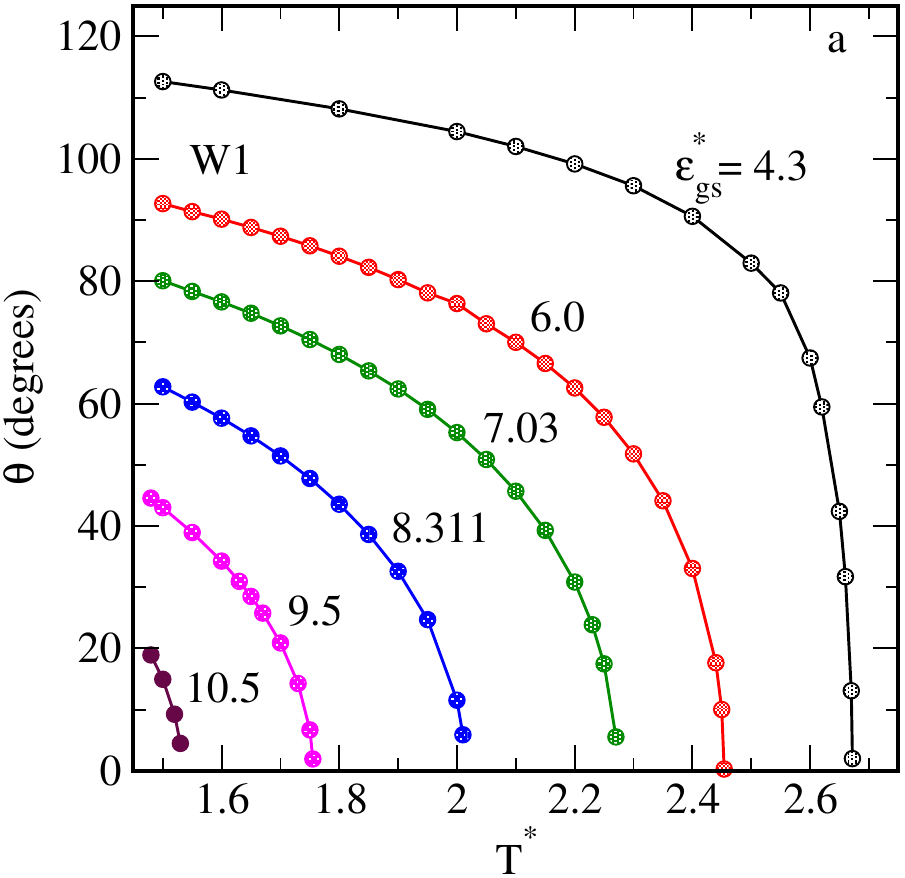}
\includegraphics[height=6.0cm]{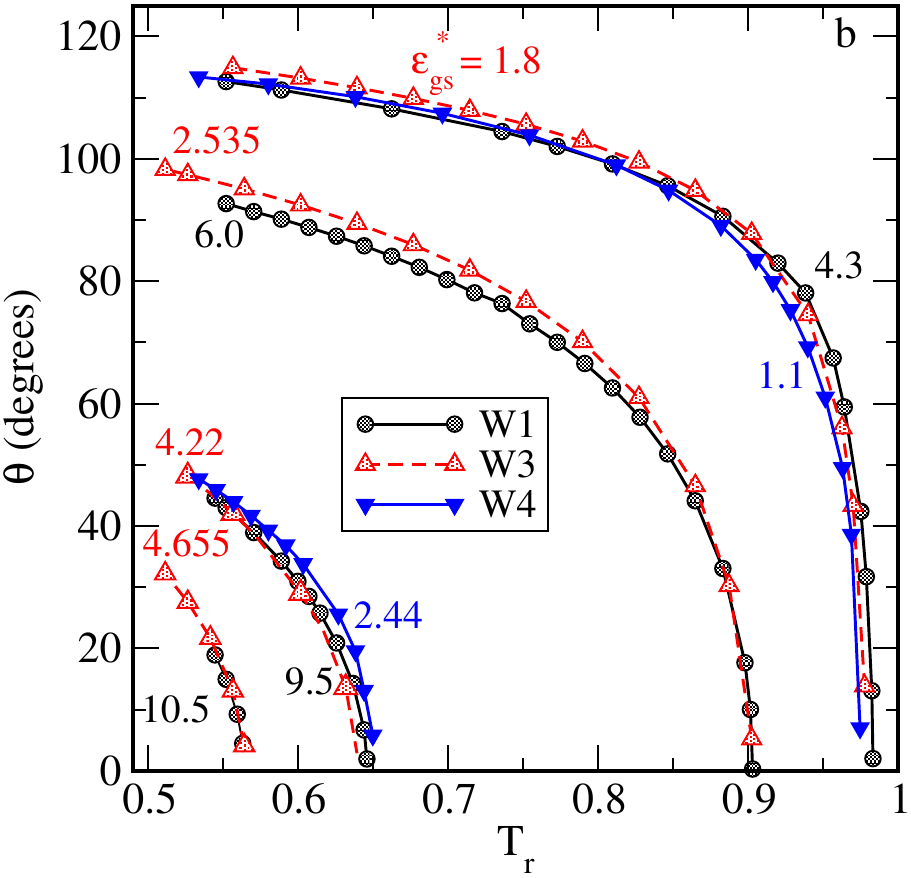}
\end{center}
\caption{(Colour online) Panel a. Temperature dependence of contact angle for W1 water-like model
in contact with various  attractive substrates. 
Panel b. A comparison of the contact angle dependence for different models W1, W3 and W4,
on rescaled temperature, $T_\text{r}=T^*/T_\text{c}^*$.
The values of the  fluid-solid attraction, $\varepsilon_{gs}^*$ (given in the figure) are chosen to
yield similar values of the wetting temperature for different models.
}
\label{fig:3}
\end{figure}

The parameter $\varepsilon_{gs}^*$ (cf. equation~(\ref{eq:egs}))
of the fluid-solid interaction is one of the principal factors
determining the adsorption of water and wettability of a given solid surface.
In figure~\ref{fig:3}a we show the values of the contact angle, $\theta$, obtained
from equation~(\ref{young}) for different values of  $\varepsilon_{gs}^*$.
Panel a is for the model W1. The temperature at which the contact angle drops
to zero is the wetting temperature, $T_w$. At temperatures higher than $T_w$,
a given surface is completely wet by liquid water, but at temperatures lower
than $T_w$ the wetting is only partial. We realize that for $\varepsilon_{gs}^*=4.3$,
the wetting temperature is very close to the critical temperature of the W1 model
(cf. table~\ref{tab:2}). Therefore, at a slightly lower value of $\varepsilon_{gs}^*$,
$\varepsilon_{gs}^*\approx 4.2$, the surface remains partially wet at all temperatures
up to the bulk critical temperature (a precise evaluation of this $\varepsilon_{gs}^*$
value is numerically tedious). For high values of $\varepsilon_{gs}^*$, the wetting
of the surface extends over a wide range of temperatures and, finally, for 
$\varepsilon_{gs}^*\gtrapprox 12$ it is
delimited from below by the bulk triple point temperature $T_\text{t}$ (the rescaled
value is $T_\text{rt}=T_\text{t}/T_\text{c}\approx 0.42$),
i.e., wetting occurs at all temperatures for which a gas can coexist with a liquid.

\begin{figure}[h]
\begin{center}
\includegraphics[height=6.0cm,clip]{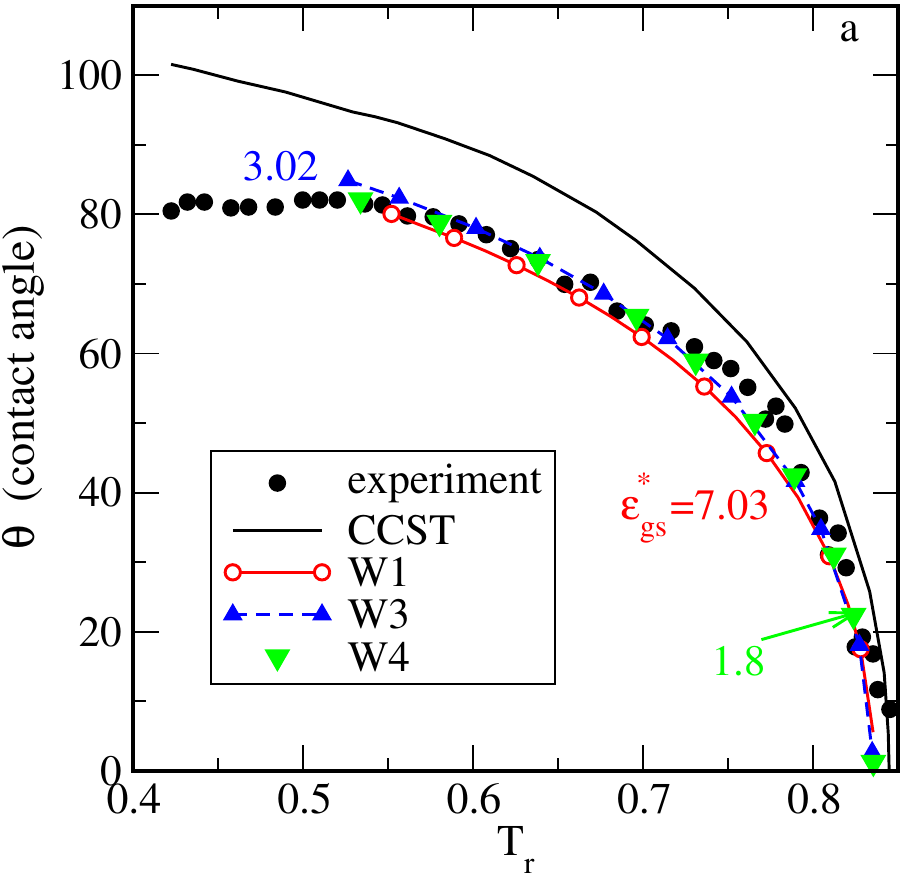}
\includegraphics[height=6.0cm,clip]{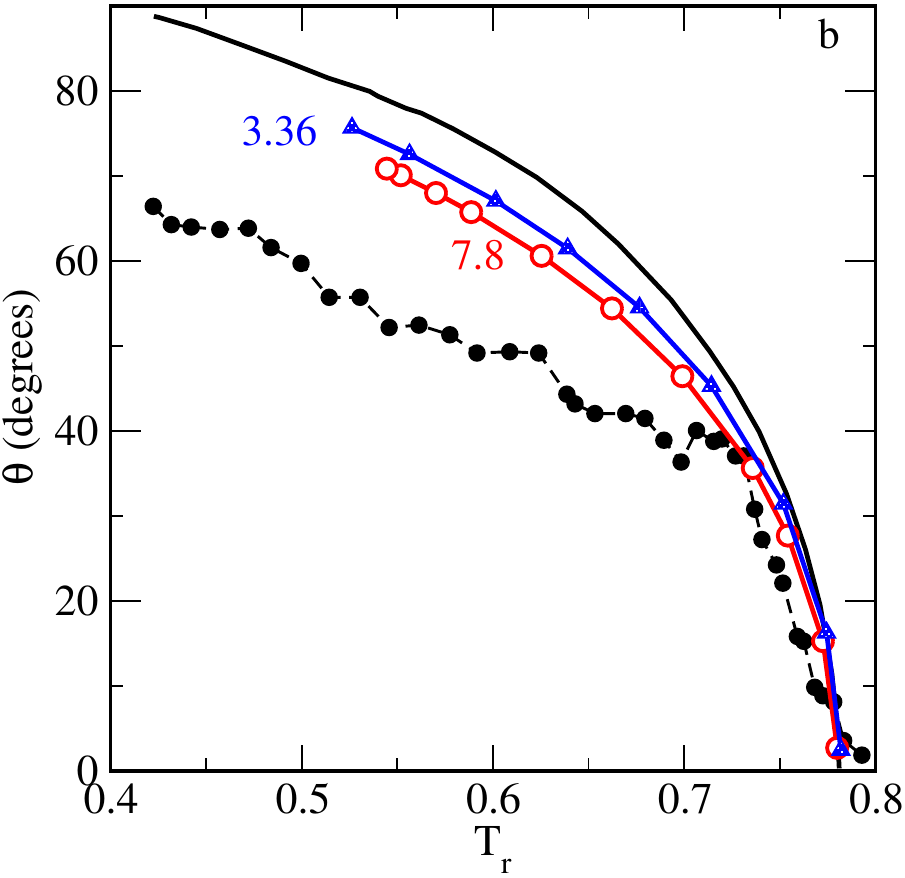}\\
\includegraphics[height=6.0cm,clip]{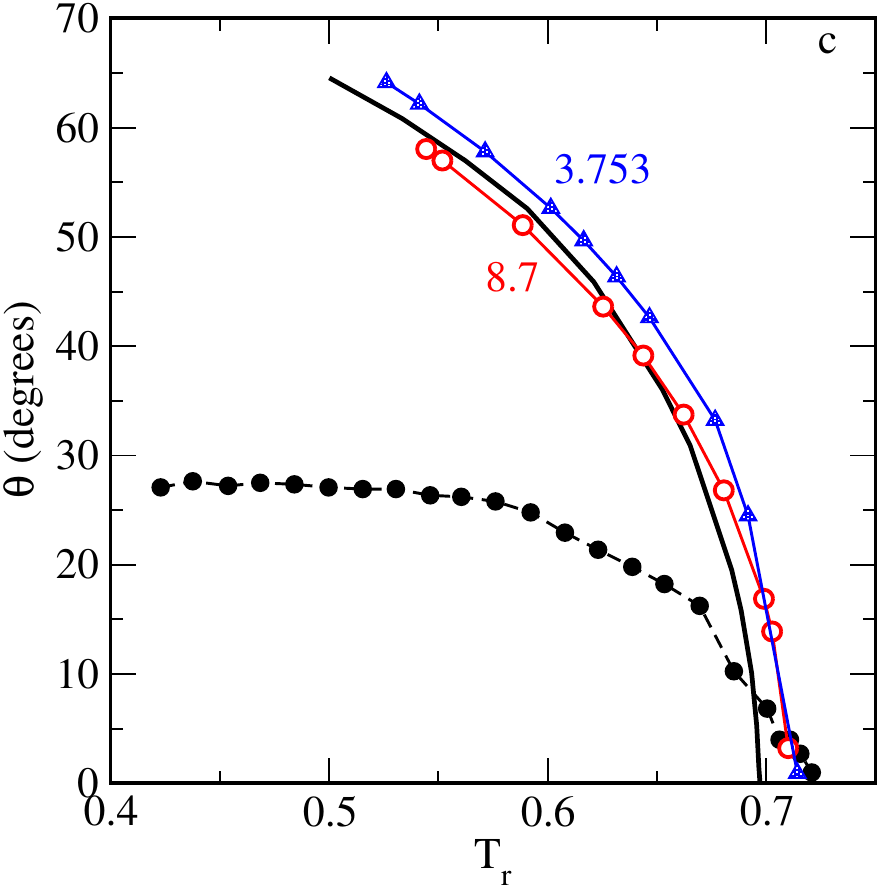}
\end{center}
\caption{(Colour online) The temperature dependence of the contact angle for W1 and W3 and W4 models
 for graphite (a), sapphire (b) and quartz (c) surfaces.  
The $\varepsilon_{gs}^*$ values 
in theoretical 
calculations are chosen to yield the ratio of the wetting temperature and the bulk
critical temperature the same as in experimental measurements \cite{es1}.
The experimental data are given by black circles. 
The solid black line in each panel
is reproduced from CCST approximation (equation~(1) of \cite{es1,cheng91}).
}
\label{fig:4}
\end{figure}

Panel b of figure~\ref{fig:3} compares the temperature dependences of $\theta$
for three models W1, W3 and W4. The temperature for each model is rescaled
as in the above ($T_\text{r}=T^*/T_\text{c}^*$). Since the values of $\varepsilon$
for each model are different, to obtain almost coinciding curves for $\theta(T_\text{r})$, different
values of $\varepsilon_{gs}^*$ should be used. They are listed in the figure.
In conclusion, we observe that a very similar temperature dependence of the contact angle
is predicted by water-like models with a differing relative weight of inter-particle 
attraction and association effects. However, the interaction potential strength 
between the water molecule and solid surface should be chosen appropriately for each model,
because the bulk behavior is different in each case.

Usually, a comparison of theoretical predictions with experimental data is not
straightforward. In particular, by considering a set of curves in figure~3b, we need
to choose the $\varepsilon_{gs}^*$ value that corresponds to the experimental
graphite. Laboratory measurements of the temperature dependence of the contact angle
of water on a highly ordered pyrolytic graphite surface are reported in~\cite{es1}.
The results indicate that the wetting temperature is at $271 \pm 12^{\circ}$C.
This estimate permits us to put the experimental data on the reduced temperature
axis together with theoretical predictions. 
In our recent study \cite{MolPhys}, we  found that for W1 model and for
 $\varepsilon_{gs}^*=7.03$, the DFT leads to almost the same 
ratio of the wetting temperature with respect to $T^*_\text{c}$ as in experiment.
For models W3 and W4, the 
values for $\varepsilon_{gs}^*$ are different and equal  $\varepsilon_{gs}^*=3.02$
and $\varepsilon_{gs}^*=1.8$, respectively. Thus, in all cases we  found ``theoretical''
graphite that corresponds to its experimental counterpart for water-like models.
The dependences of the contact angle on rescaled temperature 
for models in question, in comparison with experimental data,
are shown in figure~\ref{fig:4}a.
Besides, for the sake of comparison, we  also displayed a theoretical curve resulting 
from the approximation proposed by Cheng, Cole,  Saam, and Treiner (CCST) \cite{cheng91}
in the same figure~\ref{fig:4}a.
This approximation was derived from the Young equation
by making drastic assumptions
concerning the gas-solid and liquid-solid interfacial tensions, the expression 
for the contact angle is given by equation~(1) of~\cite{es1}.
From the inspection of all the curves, it follows that our theoretical approach 
provides an excellent description of the behavior of the contact angle in
a rather wide temperature interval. Independent of the water-water interaction
model, the results are quite satisfactory.
On the other hand, the CCST approximation is 
not appropriate in this aspect. It overestimates the contact angle values in the
entire temperature interval under study.

Similar calculations were carried out for water on sapphire, figure~\ref{fig:4}b and
on quartz,  figure~\ref{fig:4}c. According to the experiment, sapphire
surface is slightly more hydrophilic compared to graphite. 
For this system, the W1 model together with
$\varepsilon_{gs}^*=7.8$ fits the rescaled experimental wetting temperature.
On the other hand, the  W3 model together with $\varepsilon_{gs}^*=3.36$
satisfy a similar criterion. A satisfactory agreement of theoretical results
for both water-like models with the experimental trends on temperature
is observed from the wetting temperature down to $\approx 0.75 T_\text{r}$. 
At lower temperatures, the theory substantially overestimates the values for the contact 
angle.
Even a more disappointing picture emerges for water on quartz.
The  $\varepsilon_{gs}^*$ values that fit the rescaled wetting temperature are 
8.7 and 3.753 for W1 and W3 models, respectively. However, the
temperature trends are entirely unsatisfactory.
Prediction coming from the CCST approximation is of the same poor quality.
Apparent explanation of this behavior seems to be the assumption that the fluid-solid
interaction, described by the potential of Steele in equation~(\ref{steele}), is not
appropriate for the description of interaction of water molecules
with substrates more hydrophilic  than graphite. This is not surprising,
because the description of quartz-water interface requires taking account
of the chemical aspects of adsorption of water apart from the physics of adsorption,
see e.g., computer simulation setup of the problem in~\cite{notman,leeuw,gaigeot}.
Then, certain ingredients of the theoretical procedure require reconsideration 
and qualitative modification in order to take account of the bonding between 
water molecules and hydroxyl groups of the surface.

To summarize this subsection, we would like to mention that the knowledge of the 
wetting temperature values for different substrates permits to 
classify them vaguely. If the wetting temperature, $T_w$, for a given fluid
is close to the bulk critical
temperature, $T_\text{c}$, one may conclude that the surface is quite hydrophobic or say
weakly adsorbing, on intuitive terms. By contrast, if the wetting temperature is
far below the critical temperature or closer to the triple point
temperature, it is reasonable to term the substrate 
as strongly adsorbing of hydrophilic. 
However, a more sound classification of 
the substrates follows from the inspection of the shape of adsorption isotherms.
Hence, in the following subsection we turn our attention to the events
observed above the wetting temperature for water-like models on various substrates.

\subsection{Water in contact with solid surfaces. Adsorption}

If the adsorption from gaseous phase on a solid surface takes place at 
$T < T_\text{c}$, then, dependent on $T$, a 
different behavior of the adsorbed film thickness can  be observed
when density approaches the bulk saturated vapor density, $\rho_\text{bg}$.
The film thickness can either diverge
or remain finite and small. The divergence of the film thickness means
that the liquid-like film spreads on the surface. On the other hand, a constant
value of the adsorbed film thickness at $\rho_\text{bg}$ means that the liquid ``beads up''
on the surface, forming drops. The first scenario corresponds to a complete wetting,
whereas the second one --- to partial wetting.
Hence, the study of wettability can be carried out by
measuring or computing the adsorption isotherms from gaseous phase.

A comprehensive classification of
different scenarios of adsorption behavior in terms of the phase diagrams 
was elaborated by Pandit et al.~\cite{pandit82}
for systems with solely dispersive interactions.
It was shown that the principal factor 
determining the shape of adsorption isotherms on
chemical potential for different temperatures
is the ratio of the absolute values of energy of fluid-substrate attraction
and energy of fluid-fluid attraction, $w$. The latter, in fact,
establishes the temperature scale in reduced units.
For systems under study, the association between molecules is crucial, although in addition to dispersion interactions.
Thus, the ratio $w$ is not sufficient to 
describe all types of the phase behavior.

It is documented that if $w$ is high,
or the substrate is termed as strongly adsorbing, an infinite
sequence of first-order transitions, called layering transitions, can 
be observed on the adsorption isotherm.
Each of the layering transitions
corresponds to the condensation  within a given fluid layer.
If the layer number tends to infinity (while the chemical potential
approaches the chemical potential of bulk liquid-vapor coexistence, $\mu_s$),
the critical temperatures for a consecutive layering transition tend to the
so-called roughening temperature, $T_\text{R}$, which is lower 
than the bulk critical temperature.  At 
temperatures between $T_\text{R}$ and $T_\text{c}$, the layering
transitions become rounded, and the isotherm diverges
upon chemical potential approaching the $\mu_s$,
see, e.g., figure~1 of~\cite{pandit82}.

For intermediately attractive substrates, or for lower values of $w$,
the gas isotherm can exhibit a single discontinuity at 
$\mu<\mu_s$ (or $\rho_\text{b}<\rho_\text{bg}$) within a certain interval of
temperatures. This discontinuity is the manifestation of prewetting transition.
In the chemical potential-temperature plane, the prewetting transition
is a line that begins at the bulk liquid-vapor coexistence at the wetting
temperature, $T_w$.  Below the wetting temperature, the adsorption remains finite at the bulk
liquid-vapor coexistence. 
The prewetting line ends up at the prewetting critical point $(\mu(T_\text{cp}),T_\text{cp})$ and
the temperature $T_\text{cp}$  is called the prewetting critical temperature.
At temperatures above  $T_\text{cp}$, the adsorbed film thickness grows continuously  
with increasing bulk density. The entire picture is given in figure~3 of ~\cite{pandit82}.
A few comments, concerning 
the shape of adsorption isotherms presented above does not, 
of course, cover all the possibilities that result 
from different ratios, $w$. We are interested, however, in these two cases
because the water-like model in question on graphite should exhibit
similar trends according to our intuition.

\begin{figure}[h!]
\centerline{\includegraphics[height=6.5cm,clip]{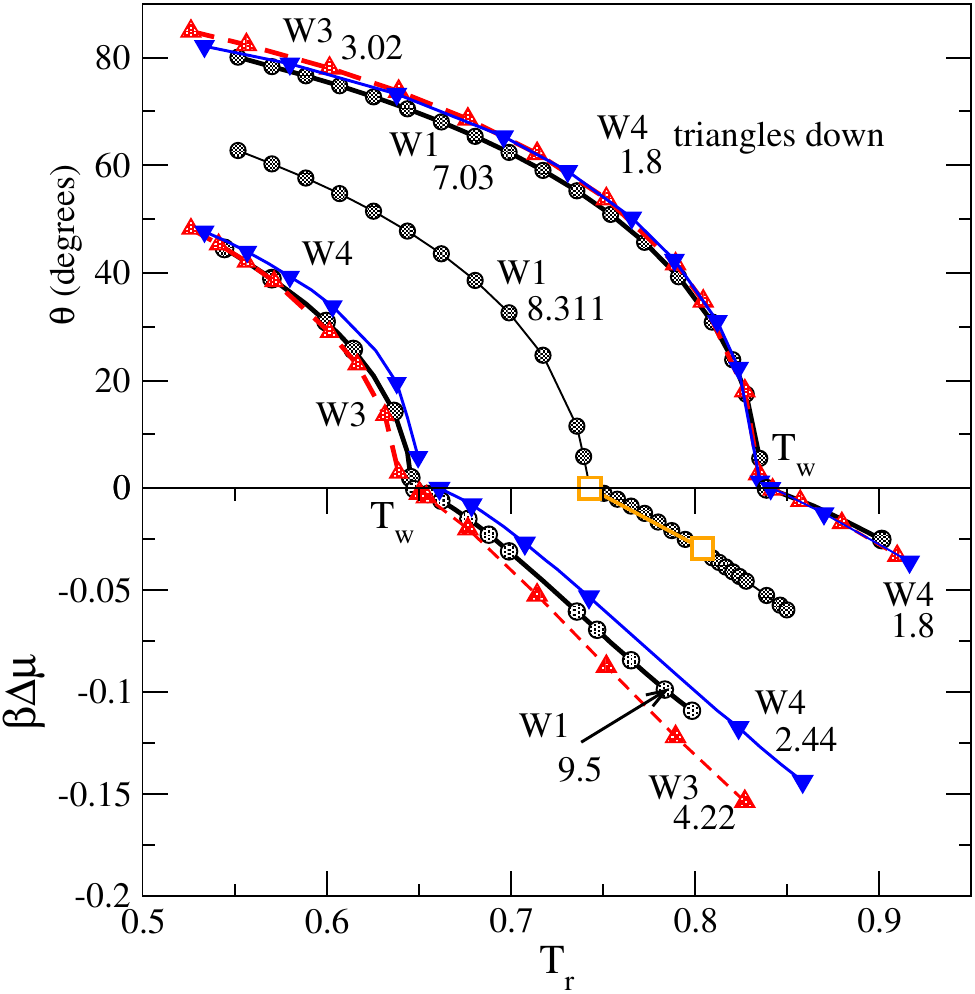}}
\caption{(Colour online) The temperature dependence of the contact
angle for W1, W3  and W4 models (upper panel) and the prewetting lines
in the temperature-chemical potential plane (lower panel). 
The values of  $\varepsilon_{gs}^*$ for each model are given in the figure.
The $X$-axis is the rescaled temperature, $T_\text{r}=T^*/T_\text{c}^*$,
cf. table \ref{tab:2}. The variable $\Delta\mu$ is defined as
$\Delta\mu = \mu -\mu_s$, where $\mu_s$ is the chemical potential at the bulk
liquid-vapor coexistence.
}
\label{fig:5}
\end{figure}

In figure~\ref{fig:5} we present the temperature dependence of
the contact angles and the corresponding estimates for the wetting temperature
(upper panel) together with the prewetting phase diagrams following from the behavior
of adsorption isotherms (lower panel) for different water-like models
and for selected values of $\varepsilon_{gs}^*$. 
The temperature axis is in rescaled units, $T_\text{r}=T^*/T_\text{c}^*$.
Both methods of determination of the wetting temperature in all cases lead to almost
identical values. On the other hand, the models with weaker association effects,
W3 and W4, predict a bit longer prewetting lines, or in other words they
yield a bit higher values for the prewetting critical temperature,
in comparison with the W1 model having a stronger association and weaker
effect of non-associative  attraction.
We should stress, however, that
the presented surface phase diagrams (lower panel) include solely
the prewetting lines. A set of layering transitions is observed only
for the W4  model at $\varepsilon_{gs}^*=2.44$, but this particular case  
will be discussed below.

We are not aware of the experimental results concerning the prewetting critical temperature.
To our best  knowledge,
the prewetting transition of water on graphite
was studied using the grand canonical ensemble computer simulation,
solely by Zhao \cite{Zhao2007}. In that study the water-water interactions 
was considered  within the SPC/E model.  
In figure ~\ref{fig:5} we compare the prewetting line resulting from simulations (squares)
with the DFT predictions for W1 model (line). In theory,
the parameter $\varepsilon_{gs}^*$ of equation~(\ref{steele}) was 
chosen equal to $\varepsilon_{gs}^*=8.311$ (this value follows from
the Lorentz-Berthelot combination rules). 
In comparison with the theory, the simulated prewetting line is remarkably shorter
indicating a different balance of association and non-association
interactions in the theory and computer simulations model.

\begin{figure}[!t]
 \begin{center}
\includegraphics[height=5.4cm,clip]{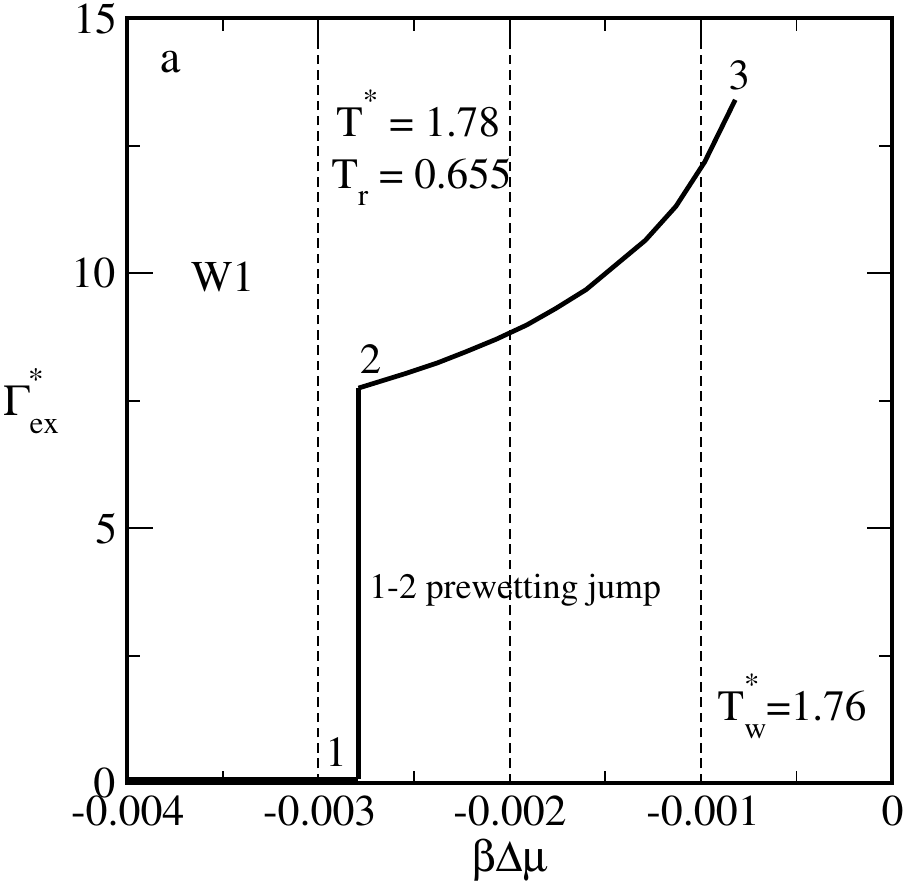}
\includegraphics[height=5.4cm,clip]{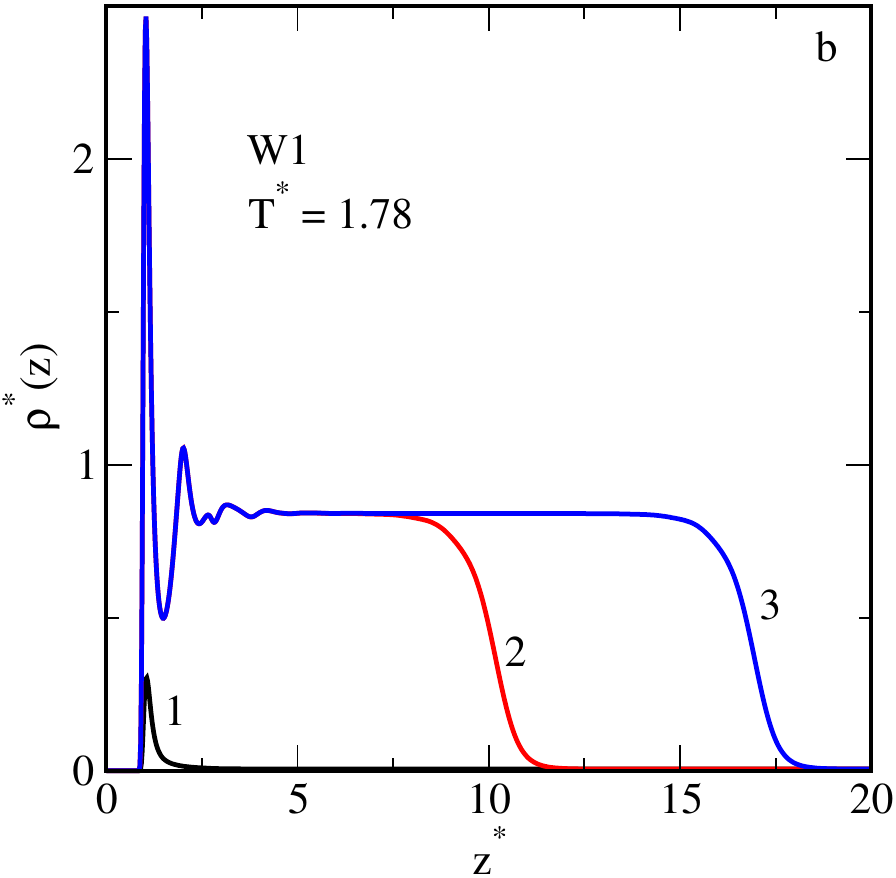}
\end{center}
\caption{(Colour online) Adsorption isotherm of water-like W1 model 
just above the wetting temperature at $T^*=1.78$ ($T_\text{r}=0.655$, $T^*_w=1.76$)
(panel a) and the density profiles of water molecules before and after the prewetting jump
and at the final state of calculations.
The labels 1 and 2 denote the prewetting jump on the adsorption isotherms.
The fluid-solid interaction is: $\varepsilon_{gs}^*=9.5$. 
}
\label{fig:6}
\end{figure}

\begin{figure}[h!]
\begin{center}
\includegraphics[height=6.0cm,clip]{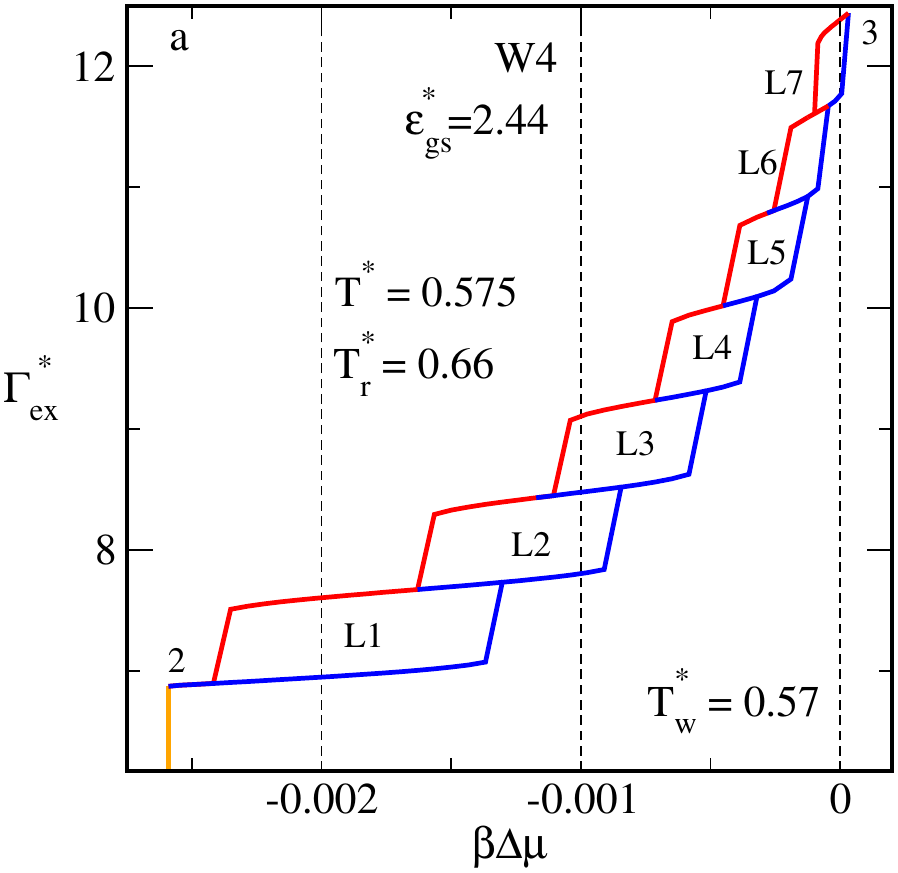}
\includegraphics[height=6.0cm,clip]{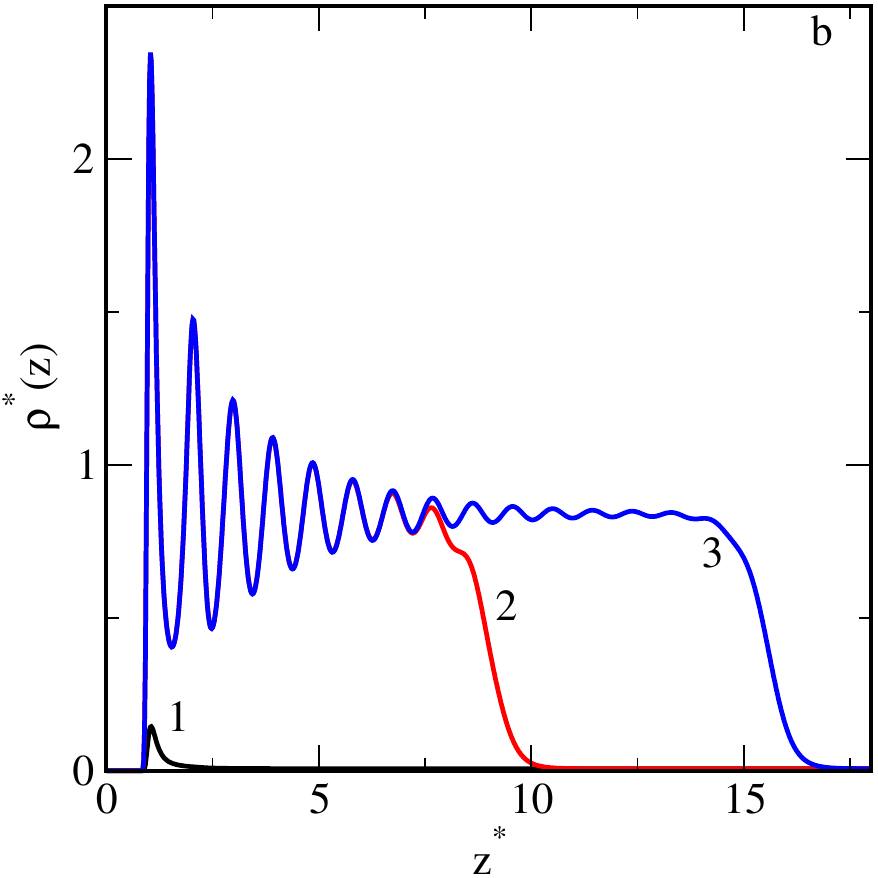}
\end{center}
\caption{(Colour online) Adsorption isotherm of water-like W4 model 
just above the wetting temperature at $T^*=0.575$ 
($T^*_w=0.57$)
(panel a) and the density profiles of water molecules before and after the prewetting jump
and at the final state of calculations.
The labels 1 and 2 denote the prewetting jump on the adsorption isotherms.
The fluid-solid interaction is $\varepsilon_{gs}^*=2.44$. 
}
\label{fig:7}
\end{figure}

The prewetting transition is manifested as a jump on the adsorption isotherm.
However, the shape of the adsorption isotherms for different water models can be different.
This issue is illustrated in figures~\ref{fig:6} and \ref{fig:7}.
The first of the above figures is for the model W1, while the second is for the model W4.
The  calculations in each case were carried out at a temperature a bit higher than the wetting
temperature, cf. figure~\ref{fig:5} ($T_\text{r} \approx 0.65$).

For  W1 model at $\varepsilon_{gs}^*=9.5$ (figure~\ref{fig:6}), the adsorption isotherm 
after the prewetting phase transition is smooth and $\Gamma_\text{ex}^*$  diverges  
as the chemical potential approaches the bulk liquid-vapor coexistence (panel a). 
The density profiles of water species at three
characteristic values of the chemical potential, marked as 1, 2 and 3 in panel a,
are shown in figure~\ref{fig:6}b.
The first value of the chemical potential  is just before and the second is just after
the prewetting jump on the adsorption isotherm, 
the third value corresponds to the highest bulk density used in the adsorption isotherm calculations.
Before the transition only a small
local density peak at the solid surface vicinity appears (figure~\ref{fig:6}b).
After the transition, the adsorbed layer extends up to the
distances $z^*\approx 10$ ($\Gamma_\text{ex}^* \approx 8$). A further increase of the chemical potential 
leads to the growth of the adsorbed film thickness, although the liquid film density remains almost constant. The thickness of the film tends to infinity as $\Delta\mu\to 0$.

 
For W4 model at $\varepsilon_{gs}^*=2.44$, however, the isotherm
consists of a series of discontinuous steps. 
Except for the first jump, the following, consecutive jumps are associated with
the condensation within single consecutive layers. 
The first jump, however, describes the formation of a thick
film on the surface. The excess adsorption increases from $\Gamma_\text{ex}^*=0.04$
(point 1, similar as in figure~6a - not shown here)
before the prewetting transition to $\Gamma_\text{ex}^* \approx 6.9$ (point 2) after it. 
The following jumps can be identified as the first-order layering transitions. 
For each transition  that is marked as L1, L2,..., L7, we plotted 
the spinodals (the equilibrium transitions appear between two spinodal parts).
The spinodals correspond to metastable adsorption and desorption branches. 
If $\Delta\mu$ tends to zero, the calculations become very tedious, since
the transitions are located very close to the bulk liquid-vapor coexistence. 
A set of seven layering transitions is observed at $T^* = 0.575$. 
The adsorption branch of the last investigated transition becomes a bit metastable
with respect to the bulk liquid-vapor transition.

The first, prewetting jump of the adsorption isotherm for W4 model
leads to the formation of the ordered film in the $OZ$-direction.
Such a situation was not observed for W1 model, i.e.,
for the model that is characterized by much stronger association effects 
than the W4 model. Thus, the layering transitions result from non-association
interactions.
Actually, the thick ordered film forms a new ``adsorbing surface''
that attracts water molecules from the gaseous phase. The adsorption
on this ``new'' surface occurs as a series of layerings. However,
the ordering in the $OZ$ direction within the consecutive layers
becomes less and less pronounced (the oscillations
of the local density profile diminish upon increasing the distance from the surface). 

\begin{figure}[h]
\centering\includegraphics[height=6.0cm,clip]{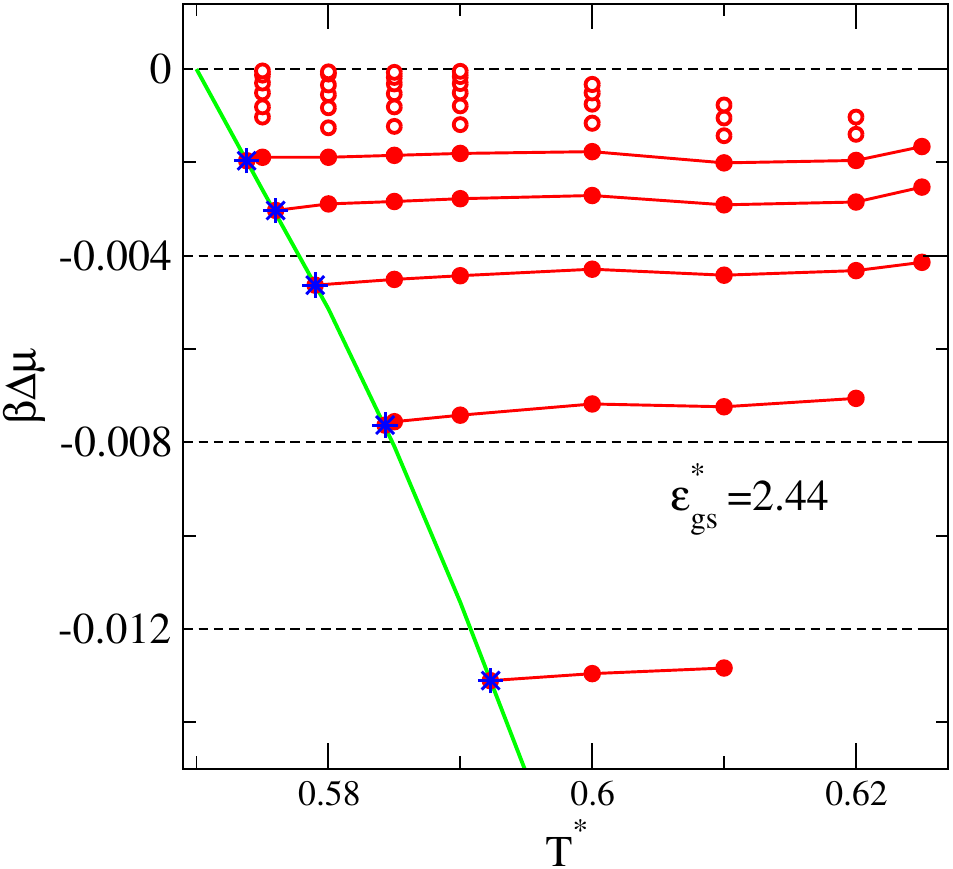}
\caption{(Colour online) The surface phase diagram in the chemical potential-temperature plane
for the model W4. The wetting temperature is $T_w^*\approx 0.57$. Only a part
of the prewetting line (green line) is plotted here. The layerings are marked
in red and the five initial triple points are marked as blue asterisks.
The fluid-solid interaction is $\varepsilon_{gs}^*=2.44$.
}
\label{fig:8}
\end{figure}

\begin{figure}[h]
\centering{\includegraphics[height=6.0cm,clip]{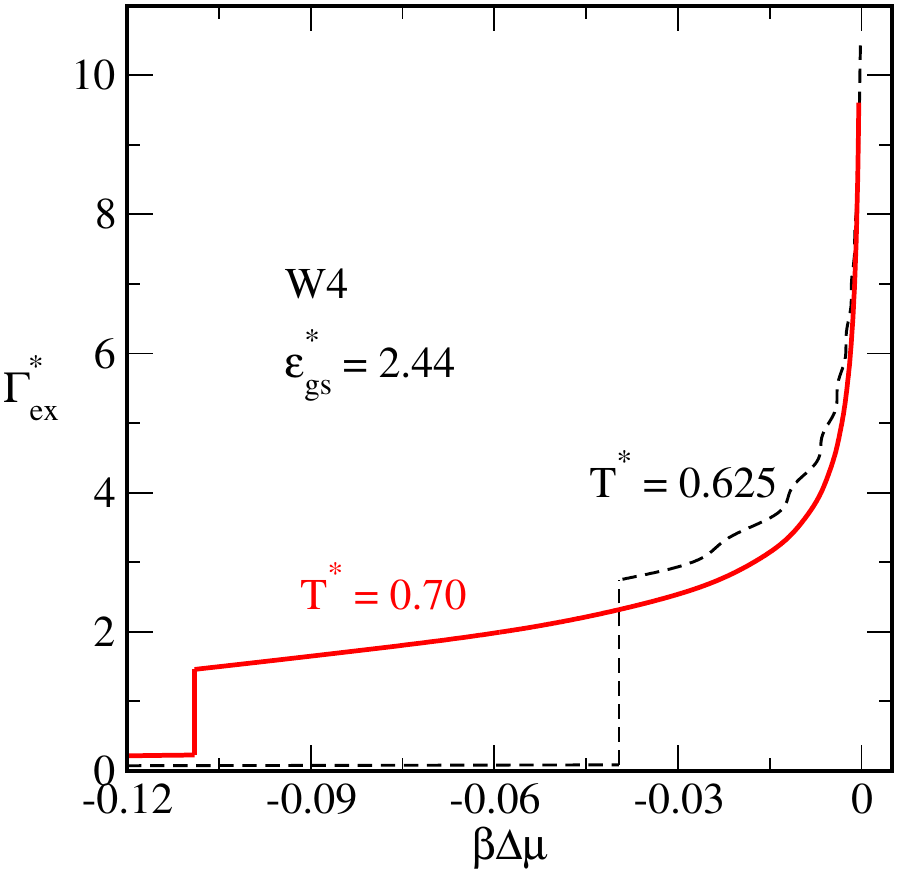}}
\caption{(Colour online) The adsorption isotherms for the W4 model at
$T^*=0.625$ (dashed line) and at $T^*=0.7$ (solid line)
}
\label{fig:9}
\end{figure}

Figure \ref{fig:8} displays a part of the surface phase diagram for the system W4.
Each layering transition line meets the prewetting line at the triple point (blue asterisk)
and ends up at its own critical point.
Complete layering diagrams are evaluated only for five  transitions
most distant from the bulk coexistence.
They are marked with solid circles. 
With an increase of the layer number, the lines for layering transitions
are getting closer to the bulk liquid-vapor coexistence and between themselves.
Thus, it becomes very  difficult to establish whether all the layerings branches
survive with decreasing temperature  till the prewetting line or 
some of them meet at a certain triple point, prior to
reaching the prewetting line, in close similarity to the study of Lennard-Jones
associating fluid in contact with solid surfaces~\cite{DF2}. The layering  
transitions 
closest to the bulk coexistence apparently end up at temperatures
above the wetting temperature as in~\cite{DF2}.
  
Definitely, the layerings L3, L4 and L5 are characterized by the higher reduced critical
point temperatures $T_{\text{{c}}Li}^*\approx 0.625$ ($i=3,4,5$) in comparison
with L1 and L2, as well as in comparison with layering transitions closest to the bulk
coexistence temperature. 
At temperatures
higher than 0.625, no layerings are present in the system and only the prewetting
transition survives, cf. figure~\ref{fig:9}. 
Indeed, at $T^*=0.625$, the isotherm is smooth, with just three steps 
layering, which are better seen in figure~\ref{fig:8}. 
The second isotherm 
in figure~\ref{fig:9} at $T^*=0.7$ exhibits solely the prewetting step. After the prewetting transition,
the isotherm is entirely smooth. During the prewetting, the adsorption increases
 from $\Gamma_\text{ex}^* \approx 0.23$ to $\Gamma_\text{ex}^* \approx 1.48$. Thus, this step is much smaller
than that observed at $T^*=0.575$ (figure~\ref{fig:7}a). 
This suggests that the temperature  0.7
is close to the surface critical temperature. 
Actually the calculations performed at $T^*=0.75$ (not shown)
lead to a typical isotherm showing a continuous increase and the divergence at the bulk gas-liquid
coexistence.  We recall (cf. figure~\ref{fig:7}) that the wetting temperature for
the system under study is $T_w^*=0.57$ and the critical prewetting temperature is close
to $T^*_\text{cp}\approx 0.74$ (see lower panel of figure~\ref{fig:5}).

The surface phase diagrams evaluated for W1 (figure~\ref{fig:5}, lower part; as well as figure~3a of
 \cite{MolPhys}) and W4 are qualitatively different. They belong to different classes
of the surface phase diagrams, according to the classification of Pandit et al. \cite{pandit82}.
However, both these diagrams are evaluated for the values of $\varepsilon_{gs}^*$ yielding a
similar value of the wetting temperature and a similar dependence  of the contact angle, $\theta$,
on temperature (in case the temperature is expressed in rescaled units),
cf. figure~\ref{fig:5}. The values of $\varepsilon_{gs}^*$ were 9.5 and 2.44 for the models W1 and W4,
respectively.

In the case of the W1  model, the entire surface phase diagram consists solely of the 
prewetting line. For the model W4, however, not only the prewetting transition but also
a sequence of layering transitions appears on the diagram. As we have already 
stressed. these models differ by the balance
of the associative and non-associative terms into the free energy functional. 
Low-dispersion, high hydrogen bonding model, W1, leads to solely prewetting type transition.
By contrast, weak chemical association and high dispersion energy between the adsorbed particles 
in the W4 model enhances 
the possibility of the development of a thick adsorbed film according 
to ``layer-by-layer'' scenario.
Otherwise, the reduction of a relative importance of the hydrogen bonds formation
causes condensation within consecutive layers, especially at lower temperatures.
However, we should emphasize that in spite of certain similarities, 
the phase diagram presented in figure~\ref{fig:8},
does not belong to any class of the surface phase diagrams discussed by Pandit et al. 
\cite{pandit82}. 
Thus, the overall picture of surface phase transitions in the
systems with associative interaction may be much richer than in the systems with
solely dispersive forces.

\section{Summary and conclusions}

In this study we revised the application of  a version of the classical density functional
approach to investigate interfacial behavior of  water at
surfaces interacting with the (10-4-3) Lennard-Jones type potential of Steele. 
Water is modelled by using SAFT methodology with
square well attraction between molecules and associative site-site interaction.
The four-site model with both dispersion and association potential energies described 
by square-well potentials with  parameters from ~\cite{Saft2} have been used.
However, in contrast to the original work of Clark et al. \cite{Saft2} 
and ~\cite{gloor},
the contribution to the nonuniform free energy functional resulting from the attractive
dispersion (non-associative) forces is approximated by the mean-field term.

The models studied by us can be considered as models ranging from predominant effects
of associative interactions compared to
non-associative effects models (W1 and W2) to the models
(W3 and W4) that are characterized by weaker association and stronger
non-associative effects.
In the case of bulk fluids, all these models lead to a reasonable description
of the bulk phase behavior of water. However, our calculations have indicated
that their application to the water-solid interfaces can lead to quite different
surface phase diagrams. In particular, the W4 model predicts a sequence of
layering transitions, while the model W1 predicts the appearance of the prewetting
line only. 

A comparison of the wetting behavior with experimental (or simulation \cite{Zhao2007})
data requires the knowledge of the parameter of the water-surface potential,
$\varepsilon_{gs}^*$. The method for this choice  used
by us is based on predicting the wetting temperature and the temperature
dependence of the contact angle. Of course, such a method does not ensure that other quantities 
such as isotherms and heat of adsorption
will also be accurately reproduced. The choice of a proper analytical
expression for the fluid-solid potential energy is a key problem
in predicting the thermodynamic properties of fluid-solid systems. This
problem is particularly well seen in the case of the temperature dependence
of the contact angle of water on quartz, and to some extend for water on sapphire. 
Despite the capability to 
correctly reproduce the wetting temperature, at lower temperatures the values of the contact angle computed
from theory significantly differ from experimental results. This may indicate the
incorrectness of the potential given by equation~(\ref{steele}) for these systems. One of possible
reasons is that the presence of hydroxyl groups on these surfaces may cause the formation
of hydrogen bonds with water. The possibility of formation of such bonds should be taken
into account upon further development of the present theory.

In order to compare the results of the present theory with experimental
data, we used the rescaling of temperature by the bulk critical temperature.
Therefore, the next improvement of the theory would relay
the construction of an approach that goes beyond the mean-field approximation.
Of course, the mean-field approximation
is computationally convenient, in contrast to any approach (e.g.,
\cite{gloor,fischer}) that takes
into account the correlations between the particles in the attractive, non-associative
free energy functional. However, all more sophisticated approaches would
result in  much more computationally demanding procedure.
Finally, it could be profitable to use a more sophisticated version of Wertheim's theory 
of association, e.g., the
second-order theory, like in \cite{kaly} for water-water hydrogen bonding and/or to take
into account water-surface associative forces (particularly for the case
of solids with hydroxyl groups on their surfaces).  Allowing for surface association,
one should also consider a possible competition between water-water and water-surface association, 
i.e., in order to modify the formulation of the mass action law.

\ukrainianpart

\title{Перегляд поведінки змочування твердих поверхонь за допомогою моделей води у рамках теорії функціоналу густини}
\author{А. Козіна\refaddr{label1}, М. Агілар\refaddr{label1}, О. Пізіо\refaddr{label1}, С. Соколовскі\refaddr{label2}}

\addresses{
	\addr{label1} Інститут хімії Національного автономного університету Мехіко, Сіркіто Екстеріор, 04510, Мехіко, Мексика
	\addr{label2} 
	Факультет теоретичної фімії, Університет ім. Марії Склодовської-Кюрі, Люблін 20-614, вул. Глиняна 33, Польща}

\makeukrtitle

\begin{abstract}
	Проведено аналіз передбачень класичної теорії функціоналу густини для асоціативних флюїдів із різною силою асоціації, що стосується змочування твердих поверхонь. Розглядаються водоподібні моделі з чотирма асоціативними силовими центрами та з неасоціативним притяганням квадратної потенціальної ями, параметризовані у роботі Кларка та ін. [Mol. Phys., {2006}, \textbf{104}, 3561]. Передбачається, що потенціал взаємодії ``рідина-тверде тіло'' має функціональну форму 10-4-3. Описано зростання водної плівки на підкладці при зміні хімічного потенціалу. Критичні температури змочування та попереднього змочування, а також фазова діаграма попереднього змочування оцінюються для різної сили притягання ``рідина-тверде тіло'' шляхом аналізу ізотерм адсорбції. Крім того, температурна залежність контактного кута отримана з рівняння Юнга; він також дає оцінки температури змочування. Теоретичні висновки порівнюються з експериментальними результатами та в кількох випадках -- з даними комп'ютерного моделювання. Запропонована теорія є досить точною щодо опису температури змочування та температурної залежності контактного кута для різних значень притягання ``рідина-субстрат''.
	Крім того, цей метод дає просте знаряддя для вивчення поведінки інших асоційованих рідин на твердих речовинах, які є важливими для хімічної інженерії, у порівнянні з лабораторними експериментами та комп’ютерним моделюванням.
	\keywords вода, графіт, функціонал густини, змочування, адсорбція
\end{abstract}

\end{document}